\documentclass[twocolumn,pre]{revtex4}

\usepackage{dcolumn}
\usepackage{amsmath}
\usepackage{calrsfs}

\usepackage{graphicx}

\newcommand{\half}{\mbox{$\frac12$}}

\newcommand{\etal}{{\it{}et~al.}}
\newcommand{\defn}{\textit}
\renewcommand{\vec}{\mathbf}
\newcommand{\mat}{\mathbf}

\newcommand\win{\omega_\textrm{in}}
\newcommand\wout{\omega_\textrm{out}}

\begin{document}

\title{Spectral methods for network community detection and graph
  partitioning}
\author{M. E. J. Newman}
\affiliation{Department of Physics and Center for the Study of Complex
Systems, University of Michigan, Ann Arbor, MI 48109}
\begin{abstract}
  We consider three distinct and well studied problems concerning network
  structure: community detection by modularity maximization, community
  detection by statistical inference, and normalized-cut graph
  partitioning.  Each of these problems can be tackled using spectral
  algorithms that make use of the eigenvectors of matrix representations of
  the network.  We show that with certain choices of the free parameters
  appearing in these spectral algorithms the algorithms for all three
  problems are, in fact, identical, and hence that, at least within the
  spectral approximations used here, there is no difference between the
  modularity- and inference-based community detection methods, or between
  either and graph partitioning.
\end{abstract}
\maketitle

\section{Introduction}
Networked systems, such as social, biological, and technological networks,
have been the subject of much recent research
activity~\cite{Newman03d,Boccaletti06}.  Along with many studies focusing
on local properties of networks, such as clustering~\cite{WS98}, degree
distributions~\cite{BA99b,ASBS00}, and correlations~\cite{PVV01,Newman02f},
are studies that examine large-scale properties like path
lengths~\cite{WS98}, percolation~\cite{CEBH00,CNSW00}, or
hierarchy~\cite{RB03,CMN08}.  Among large-scale network properties,
however, the one attracting by far the most attention has been community
structure~\cite{Fortunato10}.  Many networks are found to possess
communities or modules, groups of nodes within which connections are
relatively dense and between which they are sparser.  Communities are of
fundamental interest in networked systems because of their functional
implications---communities in a social network, for instance, may indicate
factions, interest groups, or social divisions; communities in a metabolic
network might correspond to functional units, cycles, or circuits that
perform certain tasks.

The detection of communities in network data is also of interest from an
algorithmic point of view.  It is a remarkably challenging and subtle task
for which a large number of approaches have been proposed.  In this paper
we examine two of the most widely used, the modularity maximization
method~\cite{Newman04a} and the method of statistical inference by maximum
likelihood~\cite{SN97b,ABFX08}.  In both of these approaches the community
detection problem is mapped to one of optimizing a given objective function
(either modularity or likelihood) over possible divisions of a network into
groups, but the resulting optimization problem is, in general, a
computationally hard one~\cite{Brandes07}, so one typically employs one of
a range of polynomial-time heuristics to find approximate optima, such as
Markov chain Monte Carlo~\cite{SN97b,GSA04,MAD05}, extremal
optimization~\cite{DA05}, or greedy algorithms~\cite{CNM04}.

In this paper we study one of the most elegant classes of heuristics for
network optimization problems, the spectral algorithms, inherently global
methods based on the eigenvectors of matrix representations of network
structure.  We show that both the maximum modularity and maximum likelihood
methods for community detection can be formulated as spectral algorithms
that rely on the eigenvectors of the so-called normalized Laplacian matrix.
We also describe a standard spectral algorithm for a third network problem,
the well-known problem of normalized-cut graph partitioning.  Our primary
finding is that the spectral algorithms for all three of these problems are
identical.  At least within the spectral approach taken here, there is no
difference between the detection of community structure using the methods
of maximum modularity and maximum likelihood, or between either and
normalized-cut graph partitioning.  The latter equivalence is of particular
interest because graph partitioning has been studied in depth for several
decades and a broad range of results both applied and theoretical have been
established, some of which can now be applied to the community detection
problem as well.

The outline of this paper is as follows.  In Sections~\ref{sec:modularity},
\ref{sec:inference} and~\ref{sec:partitioning} we derive in turn our
spectral algorithms for the maximum modularity, maximum likelihood, and
normalized-cut partitioning problems, which, as we have said, turn out all
to be the same.  In Section~\ref{sec:examples} we give a selection of
applications of the method to example networks, including both
computer-generated benchmark networks and real-world networks,
demonstrating its efficacy in community detection.  In
Section~\ref{sec:conclusions} we give our conclusions.

\section{Modularity maximization}
\label{sec:modularity}
In its most basic form, the problem of community detection in networks is
one of dividing the vertices of a given network into nonoverlapping groups
such that connections within groups are relatively dense while those
between groups are sparse.  As it stands, this definition is imprecise and
leaves room for interpretation, and there have, as a result, been a large
number of different methods proposed for solving the
problem~\cite{Fortunato10}.  Of these, however, probably the most widely
used is the method of modularity maximization, in which the objective
function known as modularity is optimized over possible divisions of the
network~\cite{Newman04a}.  The modularity for a given division of a network
is defined to be the fraction of edges within groups minus the expected
fraction of such edges in a randomized null model of the network.  Various
null models have been used, but the most common by far is the so-called
configuration model~\cite{MR95,NSW01}, a random graph model in which the
degrees of vertices are fixed to match those of the observed network but
edges are in other respects placed at random.  The expected number of edges
falling between two vertices $i$ and $j$ in the configuration model is
equal to $k_ik_j/2m$, where $k_i$ is the degree of vertex~$i$ and $m$ is
the total number of edges in the observed network.  The actual number of
edges observed to fall between the same two vertices is equal to the
element~$A_{ij}$ of the adjacency matrix~$\mat{A}$, so that the
actual-minus-expected edge count for the vertex pair is $A_{ij}-k_ik_j/2m$.
Giving integer labels to the groups in the proposed network division and
denoting by $g_i$ the label of the group to which vertex~$i$ belongs, the
modularity~$Q$ is then equal to
\begin{equation}
Q = {1\over2m}
    \sum_{ij} \biggl[ A_{ij} - {k_ik_j\over2m} \biggr] \delta_{g_ig_j},
\label{eq:modularity1}
\end{equation}
where $\delta_{ij}$ is the Kronecker delta.  The leading constant~$1/2m$ is
purely conventional; it has no effect on the position of the modularity
maximum.

The modularity can be calculated for divisions of a network into any number
of groups, but for the purposes of this paper we will focus on the simplest
case of division into just two groups, which is probably the most widely
studied case.

Consider, then, a network of $n$~vertices and $m$~edges, which is to be
divided into two groups of any size so as to maximize the modularity,
Eq.~\eqref{eq:modularity1}.  The modularity can be conveniently rewritten
in terms of a set of $n$ ``Ising spin'' variables~$s_i$, one for each
vertex, having values
\begin{equation}
s_i = \biggl\lbrace\begin{array}{ll}
        +1 & \quad\mbox{if vertex $i$ belongs to group~1,} \\
        -1 & \quad\mbox{if vertex $i$ belongs to group~2.}
      \end{array}
\label{eq:indices}
\end{equation}
Then $\delta_{g_ig_j} = \half (s_i s_j + 1)$ and
\begin{equation}
Q = {1\over4m} \sum_{ij} \biggl[ A_{ij} - {k_ik_j\over2m} \biggr] (s_is_j+1).
\label{eq:modularity2}
\end{equation}
We define the quantity
\begin{equation}
B_{ij} = A_{ij} - {k_ik_j\over2m}
\label{eq:bij}
\end{equation}
to be an element of a symmetric~$n\times n$ matrix~$\mat{B}$, called the
modularity matrix~\cite{Newman06b}.  The modularity matrix has the crucial
property that the sums of all its rows and columns are zero:
\begin{equation}
\sum_j B_{ij} = \sum_j A_{ij} - {k_i\over2m} \sum_j k_j
              = k_i - {k_i\over2m} 2m = 0,
\label{eq:sumrule}
\end{equation}
where we have made use of $\sum_j A_{ij} = k_i$ and $\sum_j k_j = 2m$.
Thus Eq.~\eqref{eq:modularity2} can be written as
\begin{equation}
Q = {1\over4m} \sum_{ij} B_{ij} (s_is_j+1)
  = {1\over4m} \sum_{ij} B_{ij} s_i s_j,
\label{eq:modularity3}
\end{equation}
the second term in the brackets vanishing because of~\eqref{eq:sumrule}.

The matrix elements~$B_{ij}$ are fixed once the network is given, while the
spins~$s_i$ represent the division of the network into groups.  Our task is
to maximize~$Q$ over the possible choices of the~$s_i$---the values of
$s_i$ that achieve the maximum indicate the optimal division of the network
into communities.  This is still a difficult computational task, known to
be NP-complete in general~\cite{Brandes07}, so the maximization is usually
performed using approximate heuristics.  In this paper we consider a
spectral optimization strategy, similar in spirit to the spectral method
proposed previously in~\cite{Newman06b}, but differing from it in one
crucial detail.

Maximization of~\eqref{eq:modularity3} is difficult because the
variables~$s_i$ are discrete-valued.  The problem can be made much easier
by relaxing the discreteness and allowing the~$s_i$ to take any real
values.  This is an approximation---we will be solving a somewhat different
problem from the one we really want to solve---but in practice it often
gives good results.  When we relax the~$s_i$, however, we must still impose
at least a minimal constraint on them to prevent them from becoming
arbitrarily large, which would make~$Q$ large but only in a trivial way
that yields no information about community structure.  Most commonly one
applies a constraint of the form~$\sum_i s_i^2 = n$, which limits any
individual~$s_i$ to the range $-\sqrt{n}\le s_i\le\sqrt{n}$ and fixes the
mean-square value at~1.

\begin{figure}
\begin{center}
\includegraphics[width=6cm]{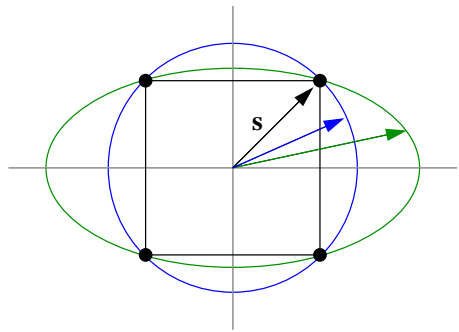}
\end{center}
\caption{Geometric representation of the relaxation method employed here.
  The true optimization is over values of the vector~$\vec{s}$ falling at
  the corner of a hypercube centered on the origin.  The most common
  relaxation involves generalizing to values that lie anywhere on the
  bounding hypersphere that touches the cube at the corners (blue).  In
  this paper, however, we generalize instead to a bounding hyperellipsoid,
  which also touches the cube at its corners (green).}
\label{fig:relax}
\end{figure}

In the language of spin models, this would be called a ``spherical
model''~\cite{BK52}.  One can think of it in geometric terms, as shown in
Fig.~\ref{fig:relax}.  If we consider the variables~$s_i$ to be the
elements of an $n$-element vector~$\vec{s}$, then the allowed
values~$s_i=\pm1$ in the original ``unrelaxed'' problem restrict the vector
to the corners of an $n$-dimensional hypercube centered on the origin,
while the relaxed values $\sum_i s_i^2 = n$ of the spherical model fall on
the bounding hypersphere of radius~$\sqrt{n}$ that touches the hypercube at
each of its corners.  Thus the relaxed values include all the allowed
values in the original problem, but also include many other values as well.

While this spherical relaxation is the commonest approach to the spectral
method, it is only one of an infinite number of possible relaxations,
differing from one another in the details of the constraint used to prevent
the values of the $s_i$ from diverging.  For instance, rather than relaxing
onto the bounding hypersphere, we can relax onto any hyperellipsoid that
touches the hypercube at all of its corners.  In other words, we can choose
a constraint of the form $\sum_i a_i s_i^2 = \sum_i a_i$ for any set of
nonnegative constants~$a_i$.  It is trivially the case that this constraint
is satisfied by the unrelaxed values~$s_i=\pm1$.  The standard hypersphere
corresponds to~$a_i=1$ for all~$i$, but in this paper we will find it
convenient to make a different choice, leading to a spectral modularity
optimization algorithm that is different in some important respects from
previous algorithms.  We set $a_i$ equal to~$k_i$, the observed degrees of
the vertices, so that our constraint takes the form
\begin{equation}
\sum_i k_i s_i^2 = 2m,
\label{eq:constraint}
\end{equation}
where $m$ is again the number of edges in the network and we have made use
of $\sum_i k_i = 2m$.

Although the original unrelaxed modularity maximization problem is a hard
one to solve, this relaxed problem is much easier.  It can be solved
exactly by simple differentiation.  Applying the
constraint~\eqref{eq:constraint} with a Lagrange multiplier~$\lambda$, the
maximum is given by
\begin{equation}
{\partial\over\partial s_l} \biggl[ \sum_{ij} B_{ij} s_i s_j
  - \lambda \sum_i k_i s_i^2 \biggr] = 0.
\end{equation}
Performing the derivatives and rearranging, we find that
\begin{equation}
\sum_j B_{ij} s_j = \lambda k_i s_i,
\label{eq:modsoln0}
\end{equation}
or, in matrix notation,
\begin{equation}
\mat{B}\vec{s} = \lambda\mat{D}\vec{s},
\label{eq:modsoln1}
\end{equation}
where $\mat{D}$ is the diagonal matrix with elements equal to the vertex
degrees~$D_{ii} = k_i$.  In other words~$\vec{s}$ is a solution of a
generalized eigenvector equation, with $\lambda$ being the eigenvalue.

To determine which eigenvector we should take, we multiply
Eq.~\eqref{eq:modsoln0} by~$s_i$ and sum over~$i$, making use
of~\eqref{eq:modularity3} and~\eqref{eq:constraint}, to get an expression
for the modularity:
\begin{equation}
Q = {1\over4m} \sum_{ij} B_{ij} s_i s_j = {\lambda\over4m} \sum_i k_i s_i^2
  = {\lambda\over2}.
\end{equation}
To achieve the highest value of the modularity, therefore, we should
choose~$\lambda$ to be the highest (most positive) eigenvalue of the
generalized eigenvector equation~\eqref{eq:modsoln1}.

Since all rows of the modularity matrix~$\mat{B}$ sum to zero, it follows
that Eq.~\eqref{eq:modsoln1} always has a solution~$\vec{s}=(1,1,1,\ldots)$
with eigenvalue~$\lambda=0$.  This solution, with all $s_i=+1$, corresponds
to putting all vertices in group~1 and none in group~2, i.e.,~not dividing
the network at all.  This tells us that if $\lambda=0$ is the highest
eigenvalue then the best modularity is achieved by not dividing the network
at all---the calculation is telling us that there is no good division of
the network into groups, so we should leave it undivided.  In our previous
work we called such networks ``indivisible.''

If, however, there is even a single strictly positive eigenvalue, then
there will exist some nontrivial solution vector~$\vec{s}$ that achieves a
higher modularity than the undivided network.  Most, though not all,
networks do have such a strictly positive eigenvalue, and we will assume
this to be the case here.

The solution above can be simplified further.  Using the
definition~\eqref{eq:bij} of the modularity matrix, we can rewrite
Eq.~\eqref{eq:modsoln0} as
\begin{equation}
\sum_j A_{ij} s_j = k_i \biggl(\lambda s_i + {1\over2m}\sum_j k_j s_j\biggr),
\end{equation}
or in matrix notation as
\begin{equation}
\mat{A}\vec{s} = \mat{D} \biggl(\lambda \vec{s} + {\vec{k}^T\vec{s}\over2m}
                 \vec{1} \biggr),
\label{eq:modsoln2}
\end{equation}
where $\vec{k}$ is the vector with elements~$k_i$ and $\vec{1} =
(1,1,1,\ldots)$.  Noting that $\mat{A}\vec{1} = \mat{D}\vec{1} = \vec{k}$
and $\vec{k}^T\vec{1}=2m$, we now multiply Eq.~\eqref{eq:modsoln2}
throughout by~$\vec{1}^T$ to get $\lambda\vec{k}^T\vec{s} = 0$, which
implies either that the largest eigenvalue~$\lambda$ is zero or that
\begin{equation}
\vec{k}^T\vec{s} = 0.
\label{eq:ortho}
\end{equation}
Since we are assuming there exists a nontrivial eigenvalue $\lambda>0$, we
know that $\lambda\ne0$ and hence~\eqref{eq:ortho} applies, which in turn
means that Eq.~\eqref{eq:modsoln2} simplifies to
\begin{equation}
\mat{A}\vec{s} = \lambda \mat{D} \vec{s}.
\label{eq:modsoln4}
\end{equation}
Thus our solution vector~$\vec{s}$ is also a solution of this generalized
eigenvector equation, involving only the standard adjacency matrix.  Again
we should choose the largest allowed value of~$\lambda$.  Now, however, the
most positive eigenvalue is disallowed---it is straightforward to see that
the uniform vector~$\vec{1}$ is an eigenvector and by the Perron-Frobenius
theorem it must have the most positive eigenvalue, since it has all
elements positive.  But this choice of eigenvector fails to satisfy
Eq.~\eqref{eq:ortho} and hence is forbidden, in which case the best we can
do is choose the eigenvector corresponding to the second most positive
eigenvalue (which can easily be shown to satisfy~\eqref{eq:ortho}, as
indeed do all the remaining eigenvectors).  This eigenvector is precisely
equal to the leading eigenvector of Eq.~\eqref{eq:modsoln1}, and hence
either~\eqref{eq:modsoln1} or~\eqref{eq:modsoln4} will give us the solution
we seek.

This is an exact solution of our relaxed modularity maximization problem.
To get a solution to the original unrelaxed problem, in which $s_i$ is
constrained to take only the values~$\pm1$, the normal approach is simply
to round the~$s_i$ to the nearest allowed value~$\pm1$.  In practice, this
just means that positive elements get rounded to~$+1$ and negative elements
to~$-1$.  Thus our final algorithm is a simple one: we calculate the
eigenvector~$\vec{s}$ of Eq.~\eqref{eq:modsoln4} corresponding to the
second-highest eigenvalue, then divide the vertices of our network into two
groups according to the signs of the elements of this vector.  This is an
approximation.  It is not guaranteed to give an exact solution to the
unrelaxed problem, but in many cases it does a good job, as we will later
see.

As a practical matter, the solution of the generalized eigenvector
equation~\eqref{eq:modsoln4} is most straightforwardly achieved by defining
a rescaled vector~$\vec{u} = \mat{D}^{1/2}\vec{s}$, where $\mat{D}^{1/2}$
is the diagonal matrix with diagonal elements equal to~$\sqrt{k_i}$.
Substituting into Eq.~\eqref{eq:modsoln4} and rearranging, we then find
that
\begin{equation}
\bigl( \mat{D}^{-1/2}\mat{A}\mat{D}^{-1/2} \bigr) \vec{u} = \lambda\vec{u}.
\end{equation}
The matrix $\mat{D}^{-1/2}\mat{A}\mat{D}^{-1/2}$ is symmetric, and thus
$\vec{u}$ is an ordinary eigenvector of a symmetric matrix, with elements
having the same signs as those of~$\vec{s}$, and with the same eigenvalue.
The spectral algorithm is thus a simple matter of calculating the
eigenvector for the second-highest eigenvalue of this symmetric matrix and
then dividing the vertices according to the signs of its elements.  For
sparse networks this can be done efficiently using sparse matrix methods
such as the Lanczos method.

The matrix
\begin{equation}
\mat{L} = \mat{D}^{-1/2}\mat{A}\mat{D}^{-1/2}
\label{eq:norml}
\end{equation}
is sometimes called the normalized Laplacian of the network and we will use
that terminology here.  (The normalized Laplacian is sometimes defined as
$\mat{I} - \mat{D}^{-1/2}\mat{A}\mat{D}^{-1/2}$, where $\mat{I}$ is the
identity, but the two matrices differ only in a trivial transformation of
their eigenvalues and the eigenvectors are the same for both.)

\section{Statistical inference}
\label{sec:inference}
We now turn to a second method for community detection in networks, the
method of statistical inference using stochastic block models.  This method
has attracted attention in recent years for the excellent results it
returns and because of the solid mathematical foundations on which it
rests, which have allowed researchers to prove rigorously a range of
results about its expected performance.  Indeed the method is provably
optimal for certain classes of networks, in the sense that no other method
will classify more vertices into their correct groups on
average~\cite{DKMZ11a,ACBL13}.

The simplest form of the method is based on the standard stochastic block
model~\cite{SN97b}, sometimes also called the planted partition
model~\cite{CK01}, a random graph model of a network containing community
structure.  The model does not itself constitute a method for community
detection.  Instead it provides a way of generating synthetic networks.  To
perform community detection, one fits the model to observed network data
using a maximum likelihood method in much the same way as one might fit a
straight line through a set of points to estimate their slope.

While formally elegant, however, this method has been found to work poorly
in practice.  The standard stochastic block model generates networks whose
vertices have a Poisson degree distribution, quite unlike the degree
distributions of most real-life networks, which means that the model is
not, typically, a good fit to observed networks for any values of its
parameters.  The situation is akin to fitting a straight line through an
inherently curved set of points---even the best fit will be a poor one
because all fits are poor.

We can get around this problem by employing a slightly more sophisticated
model, the degree-corrected block model~\cite{KN11a}, which incorporates
additional parameters that allow the model to fit non-Poisson degree
distributions, improving the fit to real-world data to the point where the
degree-corrected model appears to give good community inference in
practical situations.

The problem of fitting the degree-corrected block model to network data by
likelihood maximization is, like the modularity maximization problem, a
computationally difficult one in general, but it too can be tackled using
approximate spectral methods, as we now describe.  Indeed, as we will see,
the spectral algorithm for the degree-corrected model is ultimately
identical to the one we derived for the maximum modularity problem in the
previous section.

In the degree-corrected block model $n$ vertices are divided into groups
and edges placed between them independently at random with probabilities
that depend on the desired degrees of the vertices and on their group
membership.  Let~$g_i$ again denote the label of the group to which
vertex~$i$ belongs.  Then between each pair~$i,j$ of vertices we place a
Poisson-distributed number of edges with mean equal to~$k_i k_j
\omega_{g_ig_j}$, where $k_i$ is the desired degree of vertex~$i$ and
$\omega_{rs}$ is a set of parameters whose values control the relative
probabilities of connections within and between groups.

If $A_{ij}$ is again an element of the adjacency matrix of the observed
network, equal to the number of edges between vertices~$i$ and~$j$, then
the probability, or likelihood, that this network was generated by the
degree-corrected stochastic block model is
\begin{equation}
L = \prod_{i<j} {(k_i k_j \omega_{g_ig_j})^{A_{ij}}\over A_{ij}!}
                \exp(-k_i k_j \omega_{g_ig_j}),
\end{equation}
where the desired degrees~$k_i$ are equal to the actual degrees of the
vertices in the observed network.  Thus the likelihood of observing the
network that we did in fact observe, assuming it was generated by this
model, depends on the assignment of the vertices to the groups.  For some
assignments the network would be highly unlikely to have occurred; for
others it is more likely.  In the maximum likelihood approach, we assume
that the best assignment of vertices to groups is the one that maximizes
the likelihood.  This again turns the community detection problem into an
optimization problem which, although hard to solve exactly, often has good
approximate solutions that can be found with relative ease.

Typically, in fact, we maximize not the likelihood itself but its
logarithm~$\mathcal{L}$, which has its maximum in the same place:
\begin{equation}
\mathcal{L}
  = \frac12 \sum_{ij} \bigl[ A_{ij} \ln \omega_{g_ig_j}
    - k_i k_j \omega_{g_ig_j} \bigr],
\label{eq:loglike}
\end{equation}
where we have switched to a sum over all $i,j$ and compensated with the
leading factor of~$\frac12$, and we have assumed that the number of
edges~$A_{ij}$ between any pair of vertices is either one or zero so that
$A_{ij}!=1$ for all $i,j$.

As in Section~\ref{sec:modularity}, we will concentrate on the simplest
case of a network with just two groups, and in addition we will assume (as
most other authors also have) that there are just two different values for
the model parameters: $\win$~for pairs of vertices that fall in the same
group and~$\wout$ for pairs in different groups, with $\win>\wout$ for
traditional community structure (so-called assortative structure).
Introducing indicator variables~$s_i=\pm1$ to denote group membership as we
did in Section~\ref{sec:modularity}, we note that
\begin{align}
\omega_{g_ig_j} &= \tfrac12 \bigl[ (\win+\wout) + s_is_j (\win-\wout)
                   \bigr], \\
\ln \omega_{g_ig_j} &= \tfrac12 \biggl[ \ln (\win\wout)
               + s_is_j \ln {\win\over\wout} \biggr].
\end{align}
Substituting these expressions into Eq.~\eqref{eq:loglike}, we then find
that
\begin{equation}
\mathcal{L} = \sum_{ij} \bigl( A_{ij} - \nu k_i k_j \bigr) s_i s_j,
\label{eq:likemod}
\end{equation}
where $\nu$ is a positive constant given by
\begin{equation}
\nu = {\win-\wout\over\ln\win-\ln\wout},
\end{equation}
and we have dropped unimportant additive and multiplicative constants,
which have no effect on the position of the likelihood maximum.

Our goal is now to maximize Eq.~\eqref{eq:likemod} with respect to the
variables~$s_i$, but there is a problem: in most cases we don't know the
values of the parameters~$\win$ and~$\wout$ and hence we don't know~$\nu$
either.  Let us, however, suppose for the moment that we do know~$\nu$ and
see where it leads us.  Equation~\eqref{eq:likemod} is closely similar in
form to the modularity of Eq.~\eqref{eq:modularity3}, the only differences
being a trivial leading constant, and the substitution of $\nu$ in the
place of~$1/2m$ when compared to the modularity matrix of
Eq.~\eqref{eq:bij}.  The similarities are sufficiently strong that we can
use the same spectral approach to maximize~\eqref{eq:likemod} as we did for
the modularity, and it turns out to give the same answer.  We relax the
variables~$s_i$, allowing them to take any real values subject only to the
elliptical constraint of Eq.~\eqref{eq:constraint}, then introduce a
Lagrange multiplier~$\lambda$ and differentiate to get
\begin{equation}
\sum_j (A_{ij} - \nu k_i k_j) s_j = \lambda k_i s_i,
\label{eq:ceig1}
\end{equation}
or, in matrix notation,
\begin{equation}
(\mat{A} - \nu \vec{k}\vec{k^T}) \vec{s} = \lambda\mat{D}\vec{s},
\label{eq:ceig2}
\end{equation}
where $\mat{D}$ is the diagonal matrix of degrees as previously.  Thus the
solution to our relaxed maximization problem is an eigenvector of the
matrix~$\mat{A}-\nu\vec{k}\vec{k}^T$ and, by the same argument as before,
we should choose the leading eigenvector.

Multiplying~\eqref{eq:ceig2} on the left by~$\vec{1}^T$ and making use of
$\mat{A}\vec{1} = \mat{D}\vec{1} = \vec{k}$ and $\vec{k}^T\vec{1}=2m$, we
get
\begin{equation}
\vec{k}^T\vec{s} - 2m\nu \vec{k}^T\vec{s} = \lambda \vec{k}^T\vec{s},
\end{equation}
which implies either that $\vec{k}^T\vec{s} = 0$ or that $\lambda = 1 -
2m\nu$.  We are not at liberty, however, to choose any of the quantities
$\lambda$, $\nu$, or~$m$, and hence cannot in general satisfy the latter
condition.  Hence we must have $\vec{k}^T\vec{s} = 0$ and
Eq.~\eqref{eq:ceig2} simplifies to
\begin{equation}
\mat{A}\vec{s} = \lambda\mat{D}\vec{s},
\end{equation}
which is identical to Eq.~\eqref{eq:modsoln4} for the maximum modularity
problem.  Note that the constant~$\nu$ has dropped out of the equation, so
the fact that its value is unknown is, after all, not a problem.

From this point onward, the argument is the same as for the maximum
modularity problem and leads to the same result, that the optimal division
of the network is given by the signs of the elements of the eigenvector of
the normalized Laplacian matrix of Eq.~\eqref{eq:norml} corresponding to
the second most positive eigenvalue.

Thus, within the spectral approximation used here, the maximum modularity
and maximum likelihood methods for community detection are functionally
identical and give identical results.

\section{Graph partitioning}
\label{sec:partitioning}
We now turn to the third of the three problems mentioned in the
introduction, the problem of normalized-cut graph partitioning, which, when
tackled using the spectral method, we will show to be identical to the
community detection problems of the previous sections.  In a previous
paper~\cite{Newman13a} we noted a mapping between maximum-likelihood
community detection and the slightly different problem of minimum-cut
partitioning, although that mapping requires an extra computational step
not required by the mapping presented here.  Connections between graph
partitioning and modularity maximization have also been noted
previously~\cite{YD10,Bolla11}, although only for modified forms of the
modularity and not for the standard modularity studied in this paper.

Traditional graph partitioning is the problem of dividing a network into a
given number of parts of given sizes such that the cut size~$R$---the
number of edges running between parts---is minimized.  In the most commonly
studied case the parts are taken to be of equal size.  In many situations,
however, one is willing to tolerate a little inequality of sizes if it
allows for a better cut.  Focusing once more on the case of division into
two parts, a standard way to achieve this kind of tolerance is to minimize
not the cut size but the \defn{ratio cut}~$R/n_1n_2$, where $R$ is again
the cut size and $n_1$ and $n_2$ are the sizes of the two groups.  The
minimization is now performed with no constraint on the group sizes, but
since $n_1n_2$ is maximized when $n_1=n_2=\half n$, the minimization still
favors equally-sized groups, but it balances this favoritism against a
desire for small cut size, and the compromise seems to work well in many
practical situations.

Another variant on the same idea, which is particularly effective for
networks that have broad degree distributions, as do many real-world
networks, is minimization of the \defn{normalized
  cut}~$R/\kappa_1\kappa_2$, where $\kappa_1$ and $\kappa_2$ are the sums
of the degrees of the vertices in the two groups.  This choice favors
divisions of the network where the groups contain equal numbers of edges,
rather than equal numbers of vertices, which is desirable in certain
applications.  It is on this normalized-cut partitioning problem that we
focus in this section.

The normalized-cut problem, like the other problems we have studied, is
hard to solve exactly, but good approximate solutions can be found using
spectral methods.  The spectral approach given here is a standard one and
is not new to this paper---see, for example, Zhang and Jordan~\cite{ZJ08}.
As before, we define index variables~$s_i$ to denote the group membership
of each vertex, but rather than the $\pm1$ values we used previously, we
define
\begin{equation}
s_i = \Biggl\lbrace\begin{array}{ll}
        \rule[-7pt]{0pt}{9pt}\phantom{-}
        \sqrt{\kappa_2/\kappa_1} & \qquad\mbox{if $i$ is in group~1,} \\
        -\sqrt{\kappa_1/\kappa_2} & \qquad\mbox{if $i$ is in group~2,}
      \end{array}
\label{eq:spart}
\end{equation}
where $\kappa_1$ and~$\kappa_2$ are again the sums of the degrees of the
vertices in each group.  Note that this means that the values denoting the
two groups change when the composition of the groups changes.

With this choice for the~$s_i$, and using our previous notations~$\vec{k}$
and $\mat{D}$ for the vector and diagonal matrix of degrees respectively,
we have
\begin{align}
\vec{k}^T\vec{s} &= \sum_i k_i s_i
  = \sqrt{\kappa_2\over\kappa_1} \sum_{i\in1} k_i
    - \sqrt{\kappa_1\over\kappa_2} \sum_{i\in2} k_i \nonumber\\
  &= \sqrt{\kappa_2\kappa_1} - \sqrt{\kappa_1\kappa_2} = 0,
\label{eq:ks}
\end{align}
and
\begin{align}
\vec{s}^T\vec{D}\vec{s} &= \sum_i k_i s_i^2
  = {\kappa_2\over\kappa_1} \sum_{i\in1} k_i
    + {\kappa_1\over\kappa_2} \sum_{i\in2} k_i
  = \kappa_2 + \kappa_1 \nonumber\\
 &= 2m,
\label{eq:sds}
\end{align}
where $m$ is the number of edges in the network as before and the notation
$i\in1$ indicates that vertex~$i$ is a member of group~1.

Note also that
\begin{equation}
s_i + \sqrt{\kappa_1\over\kappa_2}
  = {2m\over\sqrt{\kappa_1\kappa_2}}\,\delta_{g_i,1},
\end{equation}
meaning this quantity is nonzero only if $i$ belongs to group~1.
Similarly
\begin{equation}
s_i - \sqrt{\kappa_2\over\kappa_1}
  = - {2m\over\sqrt{\kappa_1\kappa_2}}\,\delta_{g_i,2} .
\end{equation}

Using these results, we have
\begin{align}
& \sum_{ij} A_{ij} \biggl( s_i + \sqrt{\kappa_1\over\kappa_2} \biggr)
  \biggl( s_j - \sqrt{\kappa_2\over\kappa_1} \biggr) \nonumber\\
  &\qquad{} = -{(2m)^2\over\kappa_1\kappa_2} \sum_{ij} A_{ij}\,
    \delta_{g_i,1}\delta_{g_j,2}
  = -{(2m)^2\over\kappa_1\kappa_2} R,
\label{eq:part1}
\end{align}
where, as before, $R$~is the cut size between the two groups.  But the
quantity on the left can also be written in matrix form as
\begin{equation}
\biggl( \vec{s} + \sqrt{\kappa_1\over\kappa_2}\vec{1} \biggr)^T \mat{A}
\biggl( \vec{s} - \sqrt{\kappa_2\over\kappa_1}\vec{1} \biggr)
  = \vec{s}^T\mat{A}\vec{s} - 2m,
\label{eq:part2}
\end{equation}
where we have made use of $\vec{k} = \mat{A}\vec{1}$,
$\vec{1}^T\mat{A}\vec{1} = 2m$, and Eq.~\eqref{eq:ks}.

Combining Eqs.~\eqref{eq:part1} and~\eqref{eq:part2}, we now have a matrix
expression for the normalized cut:
\begin{equation}
{R\over\kappa_1\kappa_2} = {2m - \vec{s}^T\mat{A}\vec{s}\over(2m)^2}.
\label{eq:normcut}
\end{equation}
Thus minimizing the normalized cut is equivalent to maximizing
$\vec{s}^T\mat{A}\vec{s}$ over choices of~$s_i$
satisfying~\eqref{eq:spart}.

This hard optimization problem is once more made easier by relaxation.  We
relax the requirement that the $s_i$ take the values in
Eq.~\eqref{eq:spart}, allowing them to take any real values subject only to
the constraints~\eqref{eq:ks} and~\eqref{eq:sds}.  The relaxed problem can
then be solved straightforwardly by introducing Lagrange multipliers
$\lambda,\mu$ for the two constraints and differentiating, which gives
\begin{equation}
\mat{A}\vec{s} = \lambda\mat{D}\vec{s} + \mu\vec{k}.
\end{equation}
Multiplying on the left by~$\vec{1}^T$ and making use of $\vec{1}^T\mat{A}
= \vec{1}^T\mat{D} = \vec{k}^T$ gives
\begin{equation}
\vec{k}^T\vec{s} = \lambda \vec{k}^T\vec{s} + 2m\mu,
\end{equation}
which implies that $\mu=0$ because of Eq.~\eqref{eq:ks}, and hence we find
once again that~$\vec{s}$ is a solution of the generalized eigenvector
equation
\begin{equation}
\mat{A}\vec{s} = \lambda\mat{D}\vec{s}.
\label{eq:partsoln}
\end{equation}

Using Eqs.~\eqref{eq:sds} and~\eqref{eq:normcut}, the optimal value of the
normalized cut is then
\begin{equation}
{R\over\kappa_1\kappa_2} = {2m - \lambda
  \vec{s}^T\mat{D}\vec{s}\over(2m)^2} = {1-\lambda\over2m},
\end{equation}
which is minimized by choosing $\lambda$ as large as possible.  The leading
eigenvalue, however, is ruled out, since its eigenvector~$\vec{1}$ fails to
satisfy Eq.~\eqref{eq:ks}, so once again our solution of the relaxed
problem is given by the eigenvector corresponding to the second largest
eigenvalue of Eq.~\eqref{eq:partsoln} (which does satisfy~\eqref{eq:ks}, as
do all the other eigenvectors).

Reversing the relaxation process is a little more complicated in this case
than in the previous cases we have studied, because the discrete values
of~$s_i$ that we are rounding to, given by Eq.~\eqref{eq:spart}, are not
constant, but depend on the composition of the groups themselves.  In
principle, the most correct way to do it is to go through every possible
division of the elements of the leading eigenvector, of which there
are~$n+1$, and find the one that gives the smallest value of the normalized
cut.  In practice, however, since we are looking for solutions with roughly
equal group sizes, the values of $\kappa_1$ and $\kappa_2$ are also roughly
equal, meaning that the discrete values of~$s_i$ are approximately~$\pm1$,
and we can usually get good solutions by rounding to these values, which is
equivalent to dividing vertices according to the signs of the vector
elements.  As we show in the next section, the divisions returned by the
method are typically insensitive to the precise threshold value at which we
divide the vector elements, so the results do not depend strongly on the
rounding strategy chosen.

With this choice, which is the most common one, the algorithm becomes the
same as the algorithms we have given for community detection, either by
modularity maximization or the method of maximum likelihood.

\section{Examples}
\label{sec:examples}
We have shown that three different problems---two-way community detection
by maximum modularity and maximum likelihood, and normalized-cut bisection
of a graph---can all be solved using the same spectral algorithm.  We
compute the leading eigenvector of the normalized Laplacian matrix,
Eq.~\eqref{eq:norml}, and divide vertices according to the signs of the
vector elements.  In this section we give some example applications of the
algorithm to both computer-generated and real-world networks.

\begin{figure}
\begin{center}
\includegraphics[width=\columnwidth]{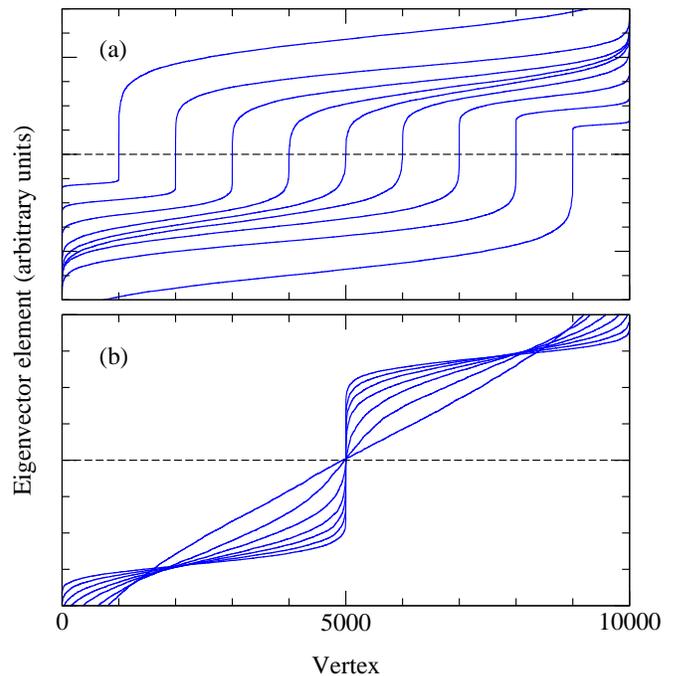}
\end{center}
\caption{Results from the application of the algorithm described here to
  networks generated using the stochastic block model with two communities.
  (a)~Each curve shows the values, plotted in increasing order, of the
  elements of the second eigenvector of Eq.~\eqref{eq:modsoln4} for single
  networks with $n=10\,000$ vertices and within-group and between-group
  edge probabilities $\win=75/n$ and $\wout=25/n$ respectively.  The curves
  are, from left to right, for networks in which group~1 has size 1000,
  2000, 3000, \dots, 9000.  (b)~Similar curves for networks of $10\,000$
  vertices and equally sized groups, but with varying edge probabilities.
  The edge probabilities are given by $n\win=60$, 65, 70, 75, 80, 85,
  and~90 and $n\wout = 100 - n\win$.}
\label{fig:sbm}
\end{figure}

Figure~\ref{fig:sbm} shows results from the application of the method to
networks generated using the stochastic block model of
Section~\ref{sec:inference}, which in addition to its use in community
inference is also widely used as a benchmark test for community detection
methods~\cite{CK01,GN02}.  Panel~(a) of the figure shows a series of curves
representing the elements of the second eigenvector of
Eq.~\eqref{eq:modsoln4} in increasing order for single networks with two
communities of varying sizes.  The horizontal dashed line indicates the
point at which the values of the elements pass zero---vertices on one side
of this line are placed in the first group and vertices on the other side
are placed in the second.  Each curve passes briskly through zero at a
point close to the sizes of the two groups planted in the network---the
size of the first group in this test was 1000, 2000, 3000, and so on for
each successive curve.  This shows that the algorithm is capable of the
accurate unsupervised detection of groups of a wide range of different
sizes.  Moreover it shows that detection is robust against
fluctuations---because the line is close to vertical as it passes zero, the
division of the network is insensitive to changes in the cut point.  If the
dashed line were moved up or down, even by quite a large amount, very few
vertices would change group membership.  This observation provides some
justification for our contention at the end of
Section~\ref{sec:partitioning} that the exact choice of the cut point is
unimportant.

Panel~(b) of the figure shows similar curves for stochastic block model
networks with two equally sized groups (which is the most challenging case)
but varying strength of community structure.  When the structure is
strongest the curves show a pronounced step at the half-way point,
indicating robust detection of the equally sized communities, but the step
becomes progressively smaller as the planted structure gets weaker, and
eventually disappears completely, so that the curve becomes featureless.
The point at which the step disappears coincides with the ``detectability
threshold'' below which it is believed that all algorithms (including this
one) must fail to detect community
structure~\cite{RL08,DKMZ11a,HRN12,NN12,MNS12}.

\begin{figure}
\begin{center}
\includegraphics[width=\columnwidth]{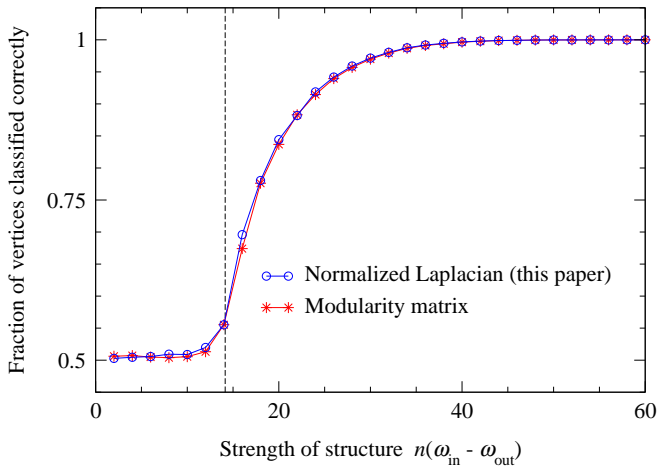}
\end{center}
\caption{The fraction of vertices classified into the correct groups by the
  algorithm described in this paper (blue circles) and by a standard
  spectral algorithm based on the leading eigenvector of the modularity
  matrix (red stars) for networks of $n=10\,000$ vertices generated from a
  stochastic block model with two equally sized groups and mean degree~50.
  The vertical dashed line represents the position of the detectability
  threshold below which all community detection algorithms must fail this
  test~\cite{RL08,DKMZ11a,HRN12,NN12,MNS12}.}
\label{fig:lb}
\end{figure}

Figure~\ref{fig:lb} further quantifies the algorithm's success at detecting
community structure in block model networks.  The figure shows the fraction
of vertices classified into the correct groups for the same situation as in
Fig.~\ref{fig:sbm}b---block model networks with $n=10\,000$ and two equally
sized groups, but varying strength of community structure.  For most of the
parameter range spanned by the figure the algorithm does a good job of
putting vertices in the right groups.  The vertical dashed line in the
figure shows the position of the detectability threshold, below which we
expect the algorithm (and indeed all algorithms) to return results no
better than a random guess (which means 50\% of vertices classified
correctly).  As we can see, the algorithm does better than random all the
way down to the transition point (if only by a small margin in the region
close to the threshold), which agrees with previous theoretical results
finding that other spectral algorithms do the same~\cite{NN12}.  Also shown
in Fig.~\ref{fig:lb} are results for tests on the same networks of the more
standard spectral community detection method of~\cite{Newman06b}, in which
one examines the leading eigenvector of the modularity matrix.  As the
figure shows, the performance of the two algorithms, at least in this test,
is essentially identical.

\begin{figure*}
\begin{center}
\includegraphics[width=1.6\columnwidth]{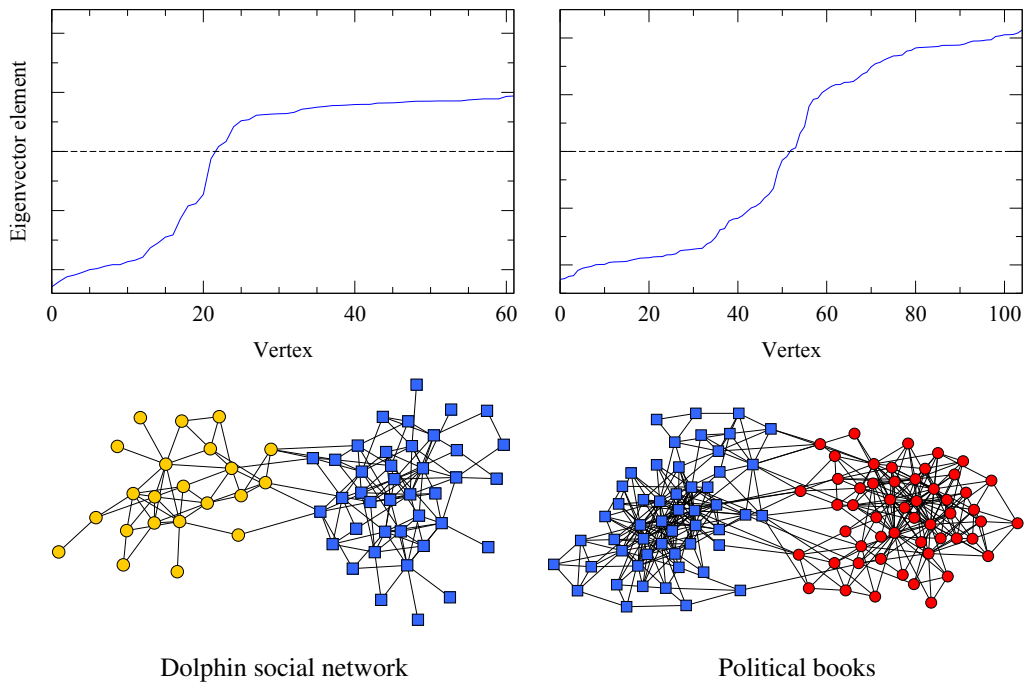}
\end{center}
\caption{Left: results of applying the algorithm to a network of frequent
  association among a group of bottlenose dolphins studied by
  Lusseau~\etal~\cite{Lusseau03a}, which is believed to divide into two
  clear communities.  The top panel shows the values of the eigenvector
  elements in increasing order.  The bottom panel shows the resulting
  division of the network.  Right: equivalent plots for a copurchasing
  network of books about US politics in which vertices represent books and
  edges connect books frequently purchased by the same purchaser.  This
  network is thought to split strongly along lines of political ideology.}
\label{fig:examples}
\end{figure*}

Figure~\ref{fig:examples} shows example applications to two well-studied
real-world networks, the dolphin social network of
Lusseau~\etal~\cite{Lusseau03a} and the political book network of
Krebs~\cite{Newman06b}, both of which are believed to break clearly into
two communities.  The top two panels in the figure show the equivalent of
the curves in Fig.~\ref{fig:sbm}---values of the elements of the second
eigenvector in increasing order.  Each shows a clear step where it crosses
the zero line (dashed lines in the plots) and the groups generated by
dividing the vertices at this point are shown in the lower panels.  In both
cases the groups correspond closely to the accepted ground truth for these
networks.

\begin{figure}
\begin{center}
\includegraphics[width=\columnwidth]{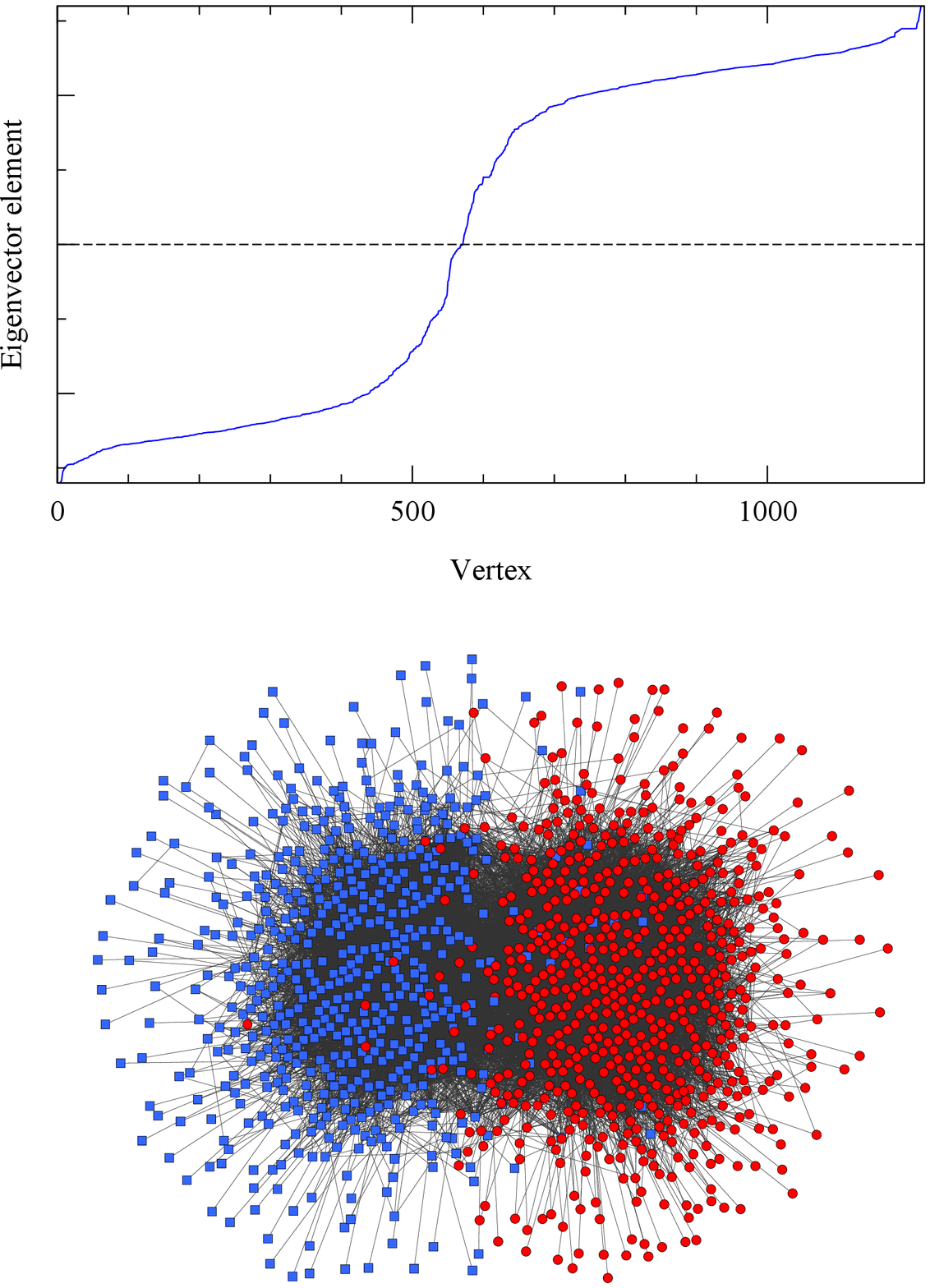}
\end{center}
\caption{Eigenvector elements and community division found by application
  of the algorithm described in this paper to a network of weblogs about US
  politics, and the web hyperlinks between them, compiled by Adamic and
  Glance~\cite{AG05}.  Again the algorithm finds clear two-way community
  structure that corresponds closely to the acknowledged division of the
  network, but in this case the community structure was obtained from the
  third, not the second, eigenvector of the normalized Laplacian.}
\label{fig:polblogs}
\end{figure}

A further interesting example is given in Fig.~\ref{fig:polblogs}, which
shows an application of the method to a network of US political weblogs
compiled by Adamic and Glance~\cite{AG05}.  Again this network is believed
to divide strongly into two communities (along lines of political outlook),
and the algorithm finds the accepted division to a good approximation.  In
this case, however, the division was found by examining the \emph{third}
eigenvector of the normalized Laplacian, not the second, as the
developments of this paper would suggest.  An examination of the second
eigenvector reveals that it is entirely uncorrelated with the community
structure in the network, instead being strongly localized around a few of
the highest-degree vertices in the network---very large vector elements for
these few hub vertices and small and apparently random elements for all
other vertices.  It is known that very high-degree vertices in networks can
give rise to high-lying, localized eigenvectors~\cite{NN13}, by mechanisms
quite different from those that produce the eigenvectors containing
community structure, and the two types of high-lying eigenvectors may
compete to be the highest in the overall spectrum.  The network of
political blogs has a particularly broad distribution of vertex degrees,
with some degrees far above the network mean, which in this case is
apparently enough to create an additional eigenvector with eigenvalue above
that of the vector containing the community structure.  Nonetheless, the
community structure is still there, clearly present in the third
eigenvector.  In practice, this means that application of the algorithm may
not be quite as simple as our derivations suggest: it may require some
finesse to extract useful community structure, particularly in the case of
networks with very high-degree hubs.  Anecdotally, based on our
experiments, we believe that the replacement of the second eigenvector with
a localized vector related to network hubs may occur more frequently in the
algorithm described in this paper than it does in more conventional
algorithms based on the eigenvectors of the modularity
matrix~\cite{Newman06b}, but this at present is merely conjecture.

\section{Conclusions}
\label{sec:conclusions}
In this paper we have given spectral algorithms for the solution of three
distinct network problems: community detection by modularity maximization,
community detection by likelihood maximization using the degree-corrected
block model, and normalized-cut graph partitioning.  As we have shown, the
algorithms for all three of these problems turn out to be the same, so that
there is no difference, at least within the spectral formulation we use,
between these three problems, although the algorithm described is different
from standard spectral algorithms for modularity maximization described in
the previous literature.  We have given results from applications of the
algorithm to a range of computer-generated and real-world networks, and it
appears to perform well in practice.

One clear possibility for extension of the calculations outlined here is
their generalization to the case of networks containing more than two
groups or communities.  The fundamental techniques needed for such a
generalization are known~\cite{ZJ08,RN13}---one replaces the index
variables~$s_i$ of Eq.~\eqref{eq:indices} with vectors pointing to the
corners of a (possibly irregular) simplex and the objective function
(modularity, likelihood, or normalized cut) with the trace of a quadratic
form involving the appropriate matrix.  At present, however, a good
all-purpose approach for community detection using such methods has yet to
be found, and so the generalization to more than two communities must be
considered an open problem.

\begin{acknowledgments}
  This work was funded in part by the National Science Foundation under
  grant DMS--1107796 and by the Air Force Office of Scientific Research
  (AFOSR) and the Defense Advanced Research Projects Agency (DARPA) under
  grant FA9550--12--1--0432.
\end{acknowledgments}

\end{document}